\documentclass{article}

\usepackage{arxiv}

\usepackage{amsmath}
\usepackage{amssymb}
\usepackage{makecell}
\usepackage[round]{natbib}

\usepackage{doi}
\makeatletter
\newcommand{\doi@}[1]{\urlstyle{same}\url{https://doi.org/#1}}
\DeclareRobustCommand{\doi}{\hyper@normalise\doi@}
\makeatother

\usepackage[utf8]{inputenc} 
\usepackage[T1]{fontenc}    
\usepackage{hyperref}       
\usepackage{url}            
\usepackage{booktabs}       
\usepackage{amsfonts}       
\usepackage{nicefrac}       
\usepackage{microtype}      
\usepackage{cleveref}       
\usepackage{lipsum}         
\usepackage{graphicx}

\title{Measuring Fractal Dimension using Discrete Global Grid Systems}

\newif\ifuniqueAffiliation

\ifuniqueAffiliation 
\author{ \href{https://orcid.org/0000-0000-0000-0000}{\includegraphics[scale=0.06]{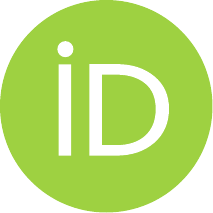}\hspace{1mm}David S.~Hippocampus}\thanks{Use footnote for providing further
		information about author (webpage, alternative
		address)---\emph{not} for acknowledging funding agencies.} \\
	Department of Computer Science\\
	Cranberry-Lemon University\\
	Pittsburgh, PA 15213 \\
	\texttt{hippo@cs.cranberry-lemon.edu} \\
	\And
	\href{https://orcid.org/0000-0000-0000-0000}{\includegraphics[scale=0.06]{orcid.pdf}\hspace{1mm}Elias D.~Striatum} \\
	Department of Electrical Engineering\\
	Mount-Sheikh University\\
	Santa Narimana, Levand \\
	\texttt{stariate@ee.mount-sheikh.edu} \\
}
\else
\usepackage{authblk}

\newbox{\orcid}\sbox{\orcid}{\includegraphics[scale=0.06]{orcid.pdf}} 
\author[1,2]{%
	\href{https://orcid.org/0000-0002-1852-5951}{\usebox{\orcid}\hspace{1mm}Pramit Ghosh\thanks{\texttt{pramitghosh@fh-muenster.de}}}%
}
\affil[1]{Innoflair UG, Richard-Wagner-Weg 35, 64287 Darmstadt, Germany}
\affil[2]{European Organisation for the Exploitation of
Meteorological Satellites (EUMETSAT), Eumetsat Allee 1, 64295 Darmstadt, Germany}
\fi

\renewcommand{\shorttitle}{Measuring Fractal Dimension using Discrete Global Grid Systems}

\hypersetup{
pdftitle={Measuring Fractal Dimension using Discrete Global Grid Systems},
pdfauthor={Pramit Ghosh},
pdfkeywords={DGGS, fractals, box-counting dimension, Hausdorff dimension, Minkowski-Bouligand dimension, MTG FCI L2 CLM},
}
\begin{document}
\maketitle

\begin{abstract}
	This study builds a bridge between two well-studied but distant topics: fractal dimension and Discrete Global Grid System (DGGS). DGGSs are used as covering sets for geospatial vector data to calculate the Minkowski-Bouligand dimension. Using the method on synthetic data yields results within 1\% of their theoretical fractal dimensions. A case study on opaque cloud fields obtained from satellite images gives fractal dimension in agreement with that available in the literature. The proposed method alleviates the problems of arbitrary grid placement and orientation, as well as the progression of cell sizes of the covering sets for geospatial data. Using DGGSs further ensure that intersections of the covering sets with the geospatial vector having large geographic extents are calculated by taking the curvature of the earth into account. This paper establishes the validity of DGGSs as covering sets theoretically and discusses desirable properties of DGGSs suitable for this purpose. 
\end{abstract}


\vspace{8mm}

\section{Introduction}
Geometrically, fractals are shapes that are invariant under similarity transformations; i.e., exhibiting self-similarity: recursive resemblance of an object to its own part – either strictly, approximately or statistically. Many geospatial phenomena in nature have been found to be fractals \citep{mandelbrot_fractal_1986}. Consequently, fractal analysis is often studied for a variety of geoscientific disciplines as listed in \citet{burrough_fractal_1981}, \citet{klinkenberg_review_1994} and \citet{goodchild_fractal_1987}. Fractals are characterised by fractal dimensions that quantifies how much space is occupied by a set near its points – a measure of the shape’s space-filling capacity. As the self-similarity of fractals in nature is only statistical and not mathematically deterministic, several methods have been developed over the years to quantify it. The box-counting method is one of the most popular ones due to its simplicity in implementation and flexibility for use with data of various kinds. Generally, it involves covering the object in question, \(F\), with tiles of size \(\delta\) and counting the minimum number of such tiles required \(N_\delta(F)\). The Minkowski-Bouligand or box-counting dimension is then given by \eqref{eq:1}. A more formal definition is given in the following section.
\[
    D_B(F)=\lim_{\delta\to0} \frac{\log N_\delta(F)}{-\log \delta}
    \tag{1}\label{eq:1}
\]
A Discrete Global Grid (DGG) is a spatial data structure that partitions the globe into areal cells and their associate cell centers \citep{sahr_geodesic_2003}. A Discrete Global Grid System (DGGS) refers to a series of DGGs with a strictly increasing number of cells. For geodesic DGGSs, spherical versions of five platonic solids (tetrahedron, hexahedron or cube, octahedron, dodecahedron and icosahedron) have been found to be the only ways to tessellate a sphere such that each cell is the same regular spherical polygon and equal number of such polygons meet at each vertex \citep{white_cartographic_2013}. As a data structure, they can be used to model both raster \citep{li_integration_2021, zheng_gpu-based_2024} and vector \citep{tong_modeling_2013, zheng_organizing_2025} data and are being used increasingly for geospatial analyses. With the DGGS OGC standard requiring interoperability of cell indices with traditional geographic coordinates \citep{purss_discrete_2017} and most implementations supporting it \citep[Table 2]{li_geospatial_2020}, earth observation data are also being processed on such grids \citep{salgues_candidate_2023, suess_processing_2004}.

Although DGGSs have been used in various applications, they have rarely been used as an operational tool or calculator, except for cellular automata models \citep{hojati_integrating_2020}. This study shows how the discrete and hierarchical nature of cells in DGGSs can be exploited to perform multiscalar covering of geospatial vector data, standardised as simple features \citep{open_geospatial_consortium_opengis_2011}, for calculating the Minkowski-Bouligand dimension. In the process, for the special case of geospatial data, relief is brought to some practical problems in calculation of the dimension that are present in the general case. The curvature of the earth that has so far been ignored while calculating Minkowski-Bouligand dimension at large geographic scales is also taken into account with the help of DGGSs.

\section{Background}
\subsection{Discrete Global Grid Systems}
Geodesic DGGSs can be defined uniquely by the choice of five independent design parameters: (1) the base regular polyhedron (popularly the icosahedron), (2) its orientation relative to the Earth, such as Fuller’s Dymaxion\textsuperscript{TM} \citep{fuller_synergetics_1975} orientation, (3) the hierarchical spatial partitioning method used (popularly hexagons, triangles and rhombi), (4) the method to transform the planar partition to its spherical equivalent, such as the Icosahedral Snyder Equal Area (ISEA) projection \citep{snyder_equal-area_1992} or the Fuller/Gray projection \citep{gray_exact_1995} and (5) the method to assign cell centers to each of the aerial cells. Different combinations of these parameters lead to the development of specific DGGSs \citep[Table 1]{sahr_geodesic_2003}.

If the increasingly finer resolution of grids of a DGGS are defined as regular planar polygons, the aperture of a DGGS is defined as the ratio of the areas of the cell at resolution \(k\) to that at resolution \(k+1\). Most DGGSs use apertures of 3, 4, 7, 9 or a combination of them.

\subsection{Fractals and their dimensions}

\subsubsection{Hausdorff Measure and Dimension}
Let the ordered pair \((M, d)\) be a metric space where \(M\) is a set and \(d\) is a metric on \(M\). Then, for a subset \(U \subset M\), its diameter \(\left|U\right|\) is defined as the greatest distance between any pair of elements in \(U\), with \(\left|\{\phi\}\right|=0\); i.e. \(\left|U\right|=\sup\{d(x, y) \mid x, y \in U\}\). If \(\{U_i\}\) be a finite collection of sets with \(0 \leq \left|U_i\right| \leq \delta\) that cover \(F \subset M\); i.e. \(F \subseteq \bigcup\limits^\infty_{i=1} U_i\), then \(\{U_i\}\) is a \(\delta\)-cover of \(F\). For any \(\delta > 0\),
\[
    H_\delta^s(F) = \inf\left\{\sum_{i=1}^\infty {\left|U_i\right|}^s \bigg| \left\{U_i\right\}\text{ is a } \delta \text{-cover of }F, s \geq 0\right\}
    \tag{2}\label{eq:2}
    \text{.}
\]
This calculates the minimum among the sums of the \(s^{\text{th}}\) powers of diameters of all the covers of \(F\) by sets of diameter at most \(\delta\). The \(s\)-dimensional Hausdorff measure of \(F\) is defined as
\[
    H^s(F) = \lim_{\delta \to 0} H_{\delta}^s(F)
    \text{.}
\]
This limit usually exists and it can be shown that it is a measure. The Hausdorff measure generalises the traditional notions of count,  length, area and volume to non-integer dimensions. For example, in \({\mathbb{R}}^n\), the Hausdorff measure is indeed proportional to the \(n\)-dimensional volume giving \(H^0(F)\) as the count of points in a set, \(H^1(F)\) as the length of a smooth curve, \(H^2(F)\) as proportional to the area of a smooth surface and so on.

The Hausdorff dimension is defined as that critical value of \(s\) for which \(H^s(F)\) jumps from infinity to 0: \(D_H(F) = \inf\left\{s \geq 0 \mid H^s(F) = 0 \right\} = \sup\left\{s \geq 0 \mid H^s(F) = \infty \right\}\). In the interval \(s \in \left[0, D_H(F)\right)\), \(H^s(F) = \infty\), while \(H^s(F) = 0\) for \(s > D_H(F)\). At  \(s=D_H(F)\), a finite value \(0 < H^s(F) < \infty\) may exist. For example, in \({\mathbb{R}}^n\), the number of points in a smooth curve given by \(H^0(F)\) is infinite, whereas its area, \(H^2(F)=0\). Only for \(s = 1\), \(H^1(F)\) has a finite non-zero value equal to the length of the curve. Thus, in this case, \(D_H(F) = 1 = D(F)\), its topological dimension.

\subsubsection{Minkowski-Bouligand Dimension}
In order to define dimension, an underlying notion is \emph{measurement at scale \(\delta\)}. It measures a set such that irregularities of size less than \(\delta\) are ignored. The Minkowski-Bouligand dimension is one such dimension among many others that uses this concept. For a non-empty bounded subset \(F \subset M\) of some metric space \(M\), if \(N_{\delta}(F)\) is the smallest number of sets that are \(\delta\)-covers of \(F\), then the Minkowski-Bouligand dimension \eqref{eq:1} is defined as the common value of the lower \eqref{eq:3} and upper \eqref{eq:4} limits if they exist and are equal \eqref{eq:5}.
\[
    \underline{D_B(F)} = \varliminf_{\delta \to 0} \frac{\log N_{\delta}(F)}{-\log \delta}
    \tag{3}\label{eq:3}
\]
\[
    \overline{D_B(F)} = \varlimsup_{\delta \to 0} \frac{\log N_{\delta}(F)}{-\log \delta}
    \tag{4}\label{eq:4}
\]
\[
    D_B(F) = \underline{D_B(F)} = \overline{D_B(F)} = \lim_{\delta \to 0} \frac{\log N_{\delta}(F)}{-\log \delta}
    \tag{5}\label{eq:5}
\]
While Hausdorff and the Minkowski-Bouligand dimensions are not equal theoretically \citep[Section 3.1]{falconer_fractal_2003}, for experimental studies they are equivalent with the later being a simplistic and computationally easier version of the former \citep{dubuc_evaluating_1989} with many regular sets yielding equal values for both the dimensions. One reason for being simpler is while in the equation for calculating \(H_{\delta}^s(F)\) a weighted sum is used for the covering sets \eqref{eq:2}, Minkowski-Bouligand dimension uses identical weights as shown in \eqref{eq:6} to calculate the smallest number of sets of diameter at most \(\delta\) that can cover \(F\).
\[
    N_{\delta}(F){\delta}^s = \inf\left\{\sum_i {\delta}^s \bigg| \left\{U_i\right\} \text{ is a finite } \delta \text{-cover of }F, s \ge 0 \right\}
    \tag{6}\label{eq:6}
\]

The formulation of this dimension as a limit of a ratio makes its calculation computationally straightforward. By plotting the minimum number of covering sets required against the reciprocal of the diameter over a suitable range of \(\delta\) on a log-log scale, the slope of the regression-line obtained through simple linear regression gives the Minkowski-Bouligand dimension. Due to ease of computation, most studies employ covering sets consisting of non-overlapping squares of varying side-lengths popularising the dimension as “box-counting” dimension even though there are no limitations on a covering set to be rectangular. Recall that the only restriction on \(\left\{U_i\right\}\) is that it must be a \(\delta\)-cover of \(F\). In this study, I use the term Minkowski-Bouligand dimension to highlight that the necessity of a covering set to be box-like or rectangular is absent.

\subsubsection{Practical Considerations}
In practice, however, computation of the Minkowski-Bouligand dimension gives rise to several practical issues. Apart from considerations from the perspective of large computer memory and availability of sufficient data points at large scale \citep{klinkenberg_review_1994}, arbitrariness in the choice of parameters while computing the dimension leads to inconsistency in results hampering reproducibility. These have to do with the covering strategy and primarily are (a) the minimum and maximum diameter of the covering sets, (b) the step size used for the diameter in the series of covering sets as \(\delta \to 0\), and (c) the position and orientation of the covering sets relative to the set whose Minkowski-Bouligand dimension is to be calculated \citep{buczkowski_modified_1998, foroutan-pour_advances_1999}. Unlike the Hausdorff dimension based on measure theory, it is well-known that the value of the Minkowski-Bouligand dimension is not absolute and depends on the parameters used. Consequently the remedies possible and available in literature are thumb rules that have given expected results on tested geometric objects usually embedded in Euclidean space covered with a set of squares \citep{liebovitch_fast_1989, theiler_estimating_1990} but have no theoretical justification in the strict sense.

It is easy to see that in the general case where the Minkowski-Bouligand dimension is being sought for an arbitrary geometric object, predefined or standardised meshes or grids that can be used as the covering sets are absent. However, in the special case of objects in the geospatial context, DGGSs provide a standardized hierarchy of such grids that are defined systematically for the entire globe that can be used as covering sets. As the aperture of DGGSs is pre-defined (and usually invariant) across resolutions and since the positions and orientations of the covering sets are fixed to the earth’s reference frame, the second and third problems in the previous paragraph are resolved respectively as they are no longer arbitrary if DGGSs are used as covering sets. This reduces the practical problems to that of choosing the range for the diameters of the cells in the covering sets. For determining the largest and the smallest cell-sizes, existing literature mostly point towards a subjective approach involving visual inspection \citep{agterberg_fractals_2013, foroutan-pour_advances_1999, ramsey_statistical_1990} although some progress has been made to find theoretical bounds on the number of cells to be sampled in the special case of square boxes in an Euclidean embedding \citep{kenkel_sample_2013, rosenberg_minimal_2016}.

\section{Methods}
I use DGGSs of types ISEA4T, ISEA4H and ISEA3H to generate covering sets at different resolution levels in the spatial extent of the spatial vector using \texttt{dggridR} \citep{barnes_dggridr_2024}. Each of the DGGSs dictate the progression used for the diameter of the cells as well as their spatial location and orientation. Maximum and minimum cell sizes of the covering sets are used so that the slope of the regression line is determined only from a limited range of scales where the data points exhibit linearity in line with accepted conventions \citep{klinkenberg_review_1994, liebovitch_fast_1989, halley_uses_2004}. I exclude resolutions for which all the individual covering cells in the bounding box of the set intersect with the set in question (thus giving the dimension of the embedding space) or the successive differences of the number of covering cells rises sharply (i.e. the number of cells covering the set do not scale linearly with the diameter of the covering set) to determine the maximum cell size. Similarly, those resolutions at which the successive differences of the number of covering cells, \(\Delta N_{\delta}\), show a sharp drop or the ratio of the number of cells of the covering set intersecting the set in question to that intersecting the bounding box of the set in question drops below 0.1 are also excluded to determine the minimum cell size.

For each resolution level, I count the number of DGGS cells intersecting with spatial vector dataset, \(N_{\delta}\) and tabulate against the reciprocal of the maximum diameter among all the intersecting DGGS cells, \({\delta}^{-1}\). Finally I run a linear regression model, \(\log N_{\delta} \sim \log {\delta}^{-1}\), to obtain a least-square regression line. The slope of the best-fit line is used as the Minkowski-Bouligand dimension, \(D_B\), of the spatial vector dataset. I also evaluate Uber H3 \citep{uber_technologies_inc_h3_2025} for suitability as covering sets through its \texttt{R} binding \texttt{h3} \citep{kuethe_h3_2022} but its high aperture of 7 results in too few data points to estimate the Minkowski-Bouligand dimension reliably from the linear regression model and, hence, I discard it.

The correctness of the values obtained for Minkowski-Bouligand dimension is checked for synthetic sets with known dimensions. Finally, I present a case study to compute the Minkowski-Bouligand dimension of a remotely sensed image of an opaque cloud field in a similar fashion to demonstrate the method’s suitability with real-world geospatial data and its generalisability.

\subsection{Sets with known Hausdorff dimensions}
In this section, with the help of two sets of known fractal dimensions represented as geospatial vector data, I establish that DGGSs can be used practically as covering sets for calculating the Minkowski-Bouligand dimension on the Earth’s surface modelled to be a sphere. Objects with large geographic extents are chosen so that the curvature of the earth is no longer trivial.

Generating the sets with \(-180 < x \le 180\) and \(-90 \le y \le 90\), where \(x\) and \(y\) are the x- and y-coordinates of the point respectively, their coordinates are treated as geographic latitude/longitude pairs and modelled as \texttt{sf} \citep{pebesma_simple_2018, pebesma_spatial_2023} objects in the \texttt{R} programming language \citep{r_lang_2024}. Algorithms to generate the geospatial vector data as a set of points are discussed in the following sub-sections. I compute the Minkowski-Bouligand dimension directly on the spatial vector dataset represented by the generated \texttt{sf} object. For the entire workflow I use spherical or great circle distances and never project the data except for the purpose of visualisation.

\subsubsection{Sierpiński Triangle}
\begin{figure}
    \centering
    \includegraphics[width=1\linewidth]{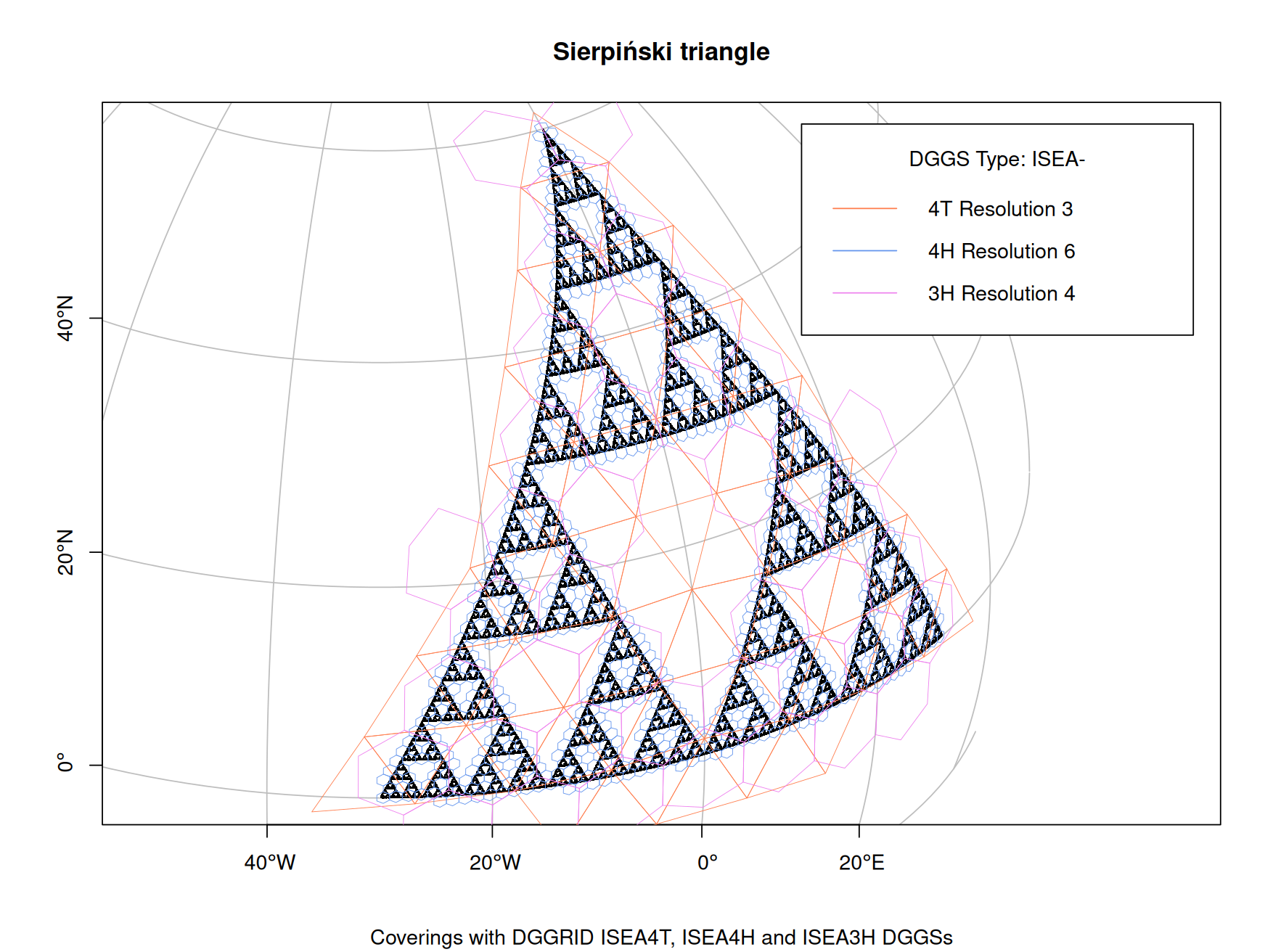}
    \caption{Embedding of the Sierpiński triangle on the globe and coverings by DGGSs}
    \label{fig:1}
\end{figure}
A Sierpiński triangle (\figureautorefname{ \ref{fig:1}}) is a fractal having the shape of a triangle and can be obtained with recursive subdivision of an equilateral triangle into four congruent parts and removing the central one. The Hausdorff dimension of the fractal is \(\frac{\log 3}{\log 2} \approx 1.585\). I generate the fractal as a chaos game using a randomised algorithm \citep[Section 17.4]{feldman_chaos_2012}:
\begin{enumerate}
    \item Choose the vertices of the external triangle as the coordinates \(A(-30,0)\), \(B(0,60)\) and \(C(30,0)\).
    \item Choose an initial current point to be at the origin \(P^{(0)}(0, 0)\).
    \item Iterate the following steps until the maximum number of iterations, \(MAX = 200000\) is reached:
    \begin{enumerate}
        \item Store \(P^{(i)}(x, y)\).
        \item Randomly choose one of the three vertices of the triangle (\(A\), \(B\) or \(C\)). Let it be \(V(x_v, y_v)\).
        \item Choose the current point for the next iteration, \(P^{(i+1)}\) to be the midpoint of the line joining the points \(P^{(i)}\) and \(V\); i.e. the coordinates \(P^{(i+1)}\left(\frac{x+x_v}{2}, \frac{y+y_v}{2}\right)\).
    \end{enumerate}
\end{enumerate}
Once the points have been generated, I transform them to a \(MAX \times 2\) matrix where the columns represent the x- and y-coordinates of the points.

\subsubsection{Koch Curve}
\begin{figure}
    \centering
    \includegraphics[width=1\linewidth]{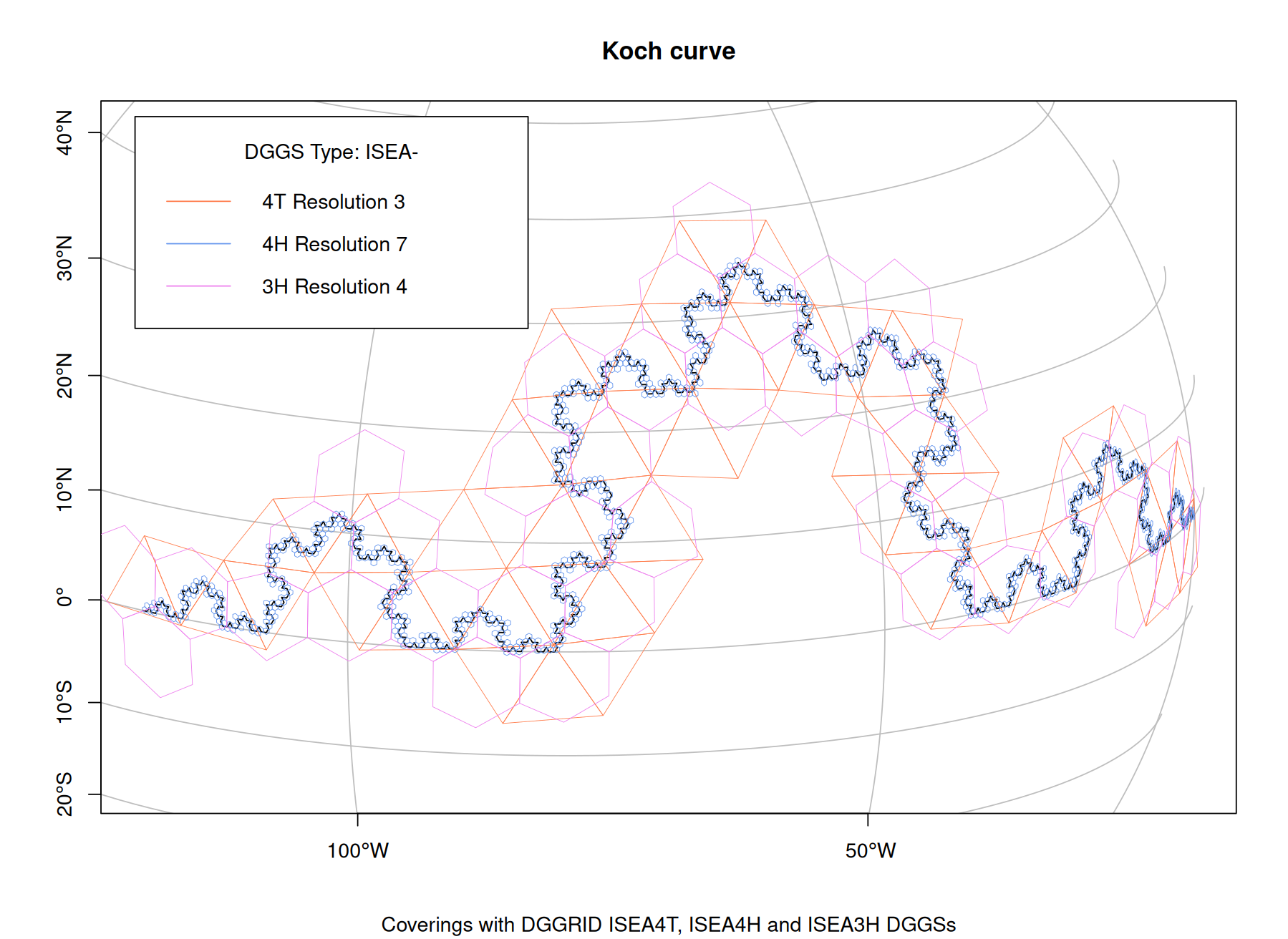}
    \caption{Embedding of the Koch curve on the globe and and coverings with DGGSs}
    \label{fig:2}
\end{figure}

The Koch curve (\figureautorefname{ \ref{fig:2}}) was initially proposed as an example of a geometric curve that is continuous everywhere but differentiable nowhere with a Hausdorff dimension of \(\frac{\log 4}{\log 3} \approx 1.261\). It can be constructed iteratively from an initial seed of a simple line segment. Every iteration involves replacing the middle third of each line segment with the two sides of an equilateral triangle – i.e. without the base. Formally, a Koch curve can be written using the Lindenmayer- or L-system as follows. Here \(F\) implies going forward while \(+\) and \(-\) imply turning anti-clockwise and clockwise respectively by the amount defined as the angle parameter.
\begin{itemize}
    \item Axiom: \(F\)
    \item Production rules: \(F \to F+F--F+F\)
    \item Angle: \(60^{\circ}\)
\end{itemize}
For the purposes of this study, I write a program that takes the L-system given above as input and by applying the production rules successively generates a string for 5 iterations. Finally, the program computes the coordinates for each point by traversing the string in a manner similar to turtle graphics. The coordinates are scaled uniformly by a factor of 0.5 so that the maximum value on the vertical axis is less than 90 (so that the coordinates can be translated easily to latitude/longitude values).

\subsection{Fractal Clouds} \label{sec:fractal_clouds}
The fractals discussed so far are ideal ones that are strictly self-similar but in nature fractals are only self-similar statistically. Clouds have been first identified to exhibit fractal behaviour by \citet{lovejoy_area-perimeter_1982} and several studies have since been done to determine the fractal dimension of the 2-dimensional projection cloud fields or ensembles spanning large geographic regions. While some variation has been noted depending on the type of clouds, reported values are mostly consistent with \(D_B \approx 1.56\) for large clouds and cloud fields \citep{cahalan_fractal_1989, lovejoy_review_2023, rees_global_2024, yano_self-similarity_1987}.

\begin{figure}
    \centering
    \includegraphics[width=1\linewidth]{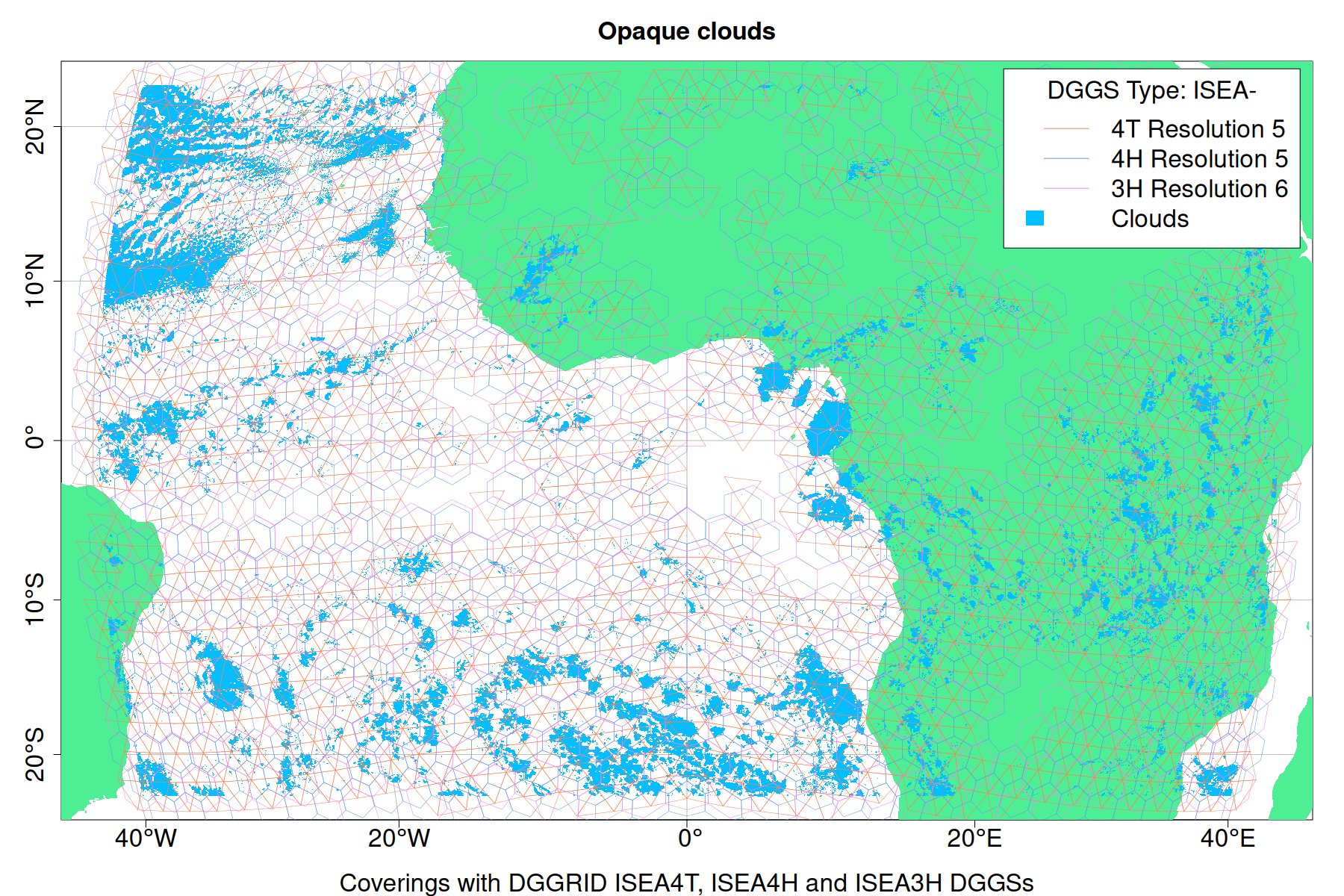}
    \caption{Opaque cloud fields obtained from MTG FCI Level 2 Cloud Mask product and coverings with different DGGSs}
    \label{fig:3}
\end{figure}

They can be observed over large geographical regions using geostationary remote sensing meteorological satellites such as the Meteosat Third Generation (MTG) satellite Meteosat-11. The MTG Level-2 Cloud Mask product \citep{eumetsat_eumetsat_2025} obtained by downstream processing of Level-1 images from the Flexible Combined Imager (FCI) instrument onboard the satellite identifies clouds among other atmospheric artefacts. I customise a cloud mask product sensed at midday of 20th March 2025 using the Data Tailor Web Service \citep{eumetsat_data_2025} with the following parameters: (a) Output format: GeoTIFF, (b) Layer: Cloud Mask (Channel 1), (c) Projection: Geographic, (d) Region of interest: 40°W-40°E, 23°S-23°N to obtain a raster image. The reasons for choosing this region is that pixels in high latitudes get distorted significantly and this region overlaps with the Intertropical Convergence Zone (ITCZ) – a band of low pressure region around the equator that frequently give rise to thunderstorms and, therefore, an important region of study from a meteorological point of view. The interaction with the web service and subsequent download of the customized product is performed using \texttt{EUMDAC} \citep{horn_eumdac_2024}. I create a binary cloud mask (\figureautorefname{ \ref{fig:3}}) using \texttt{terra} \citep{hijmans_terra_2025} by selecting only those pixels that represent opaque clouds (\(DN = 3\)) and subsequently polygonise it so that contiguous pixels representing a cloud are clumped into a polygon.

Once the clouds are represented by a set of polygons, I tabulate the cell diameters and the number of cells required to cover the clouds. I use a linear regression model to find the best-fit line and calculate the Minkowski-Bouligand dimension in the manner stated previously.

\section{Results}
\begin{figure}
    \centering
    \includegraphics[width=1\linewidth]{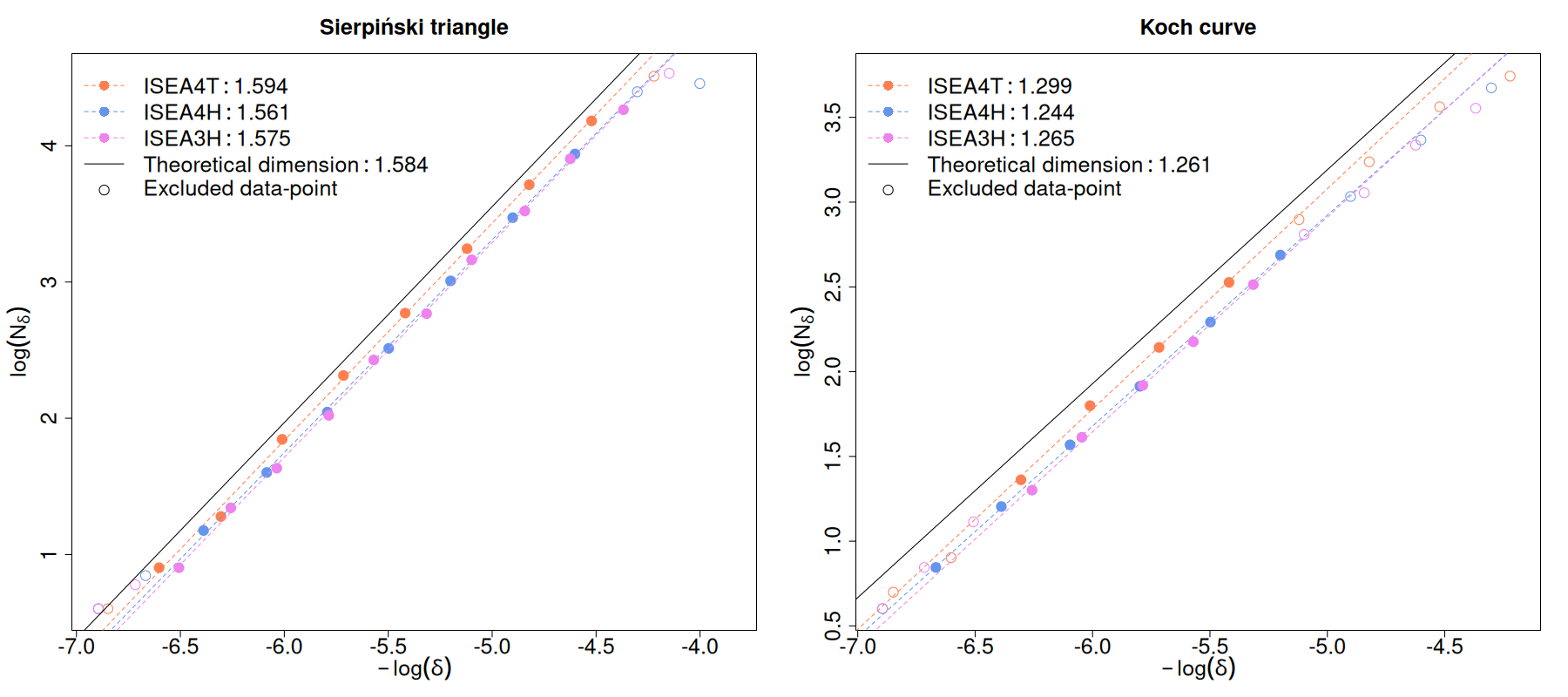}
    \caption{Calculation of the Minkowski-Bouligand dimensions of the Sierpiński Triangle (left) and the Koch Curve (right)}
    \label{fig:4}
\end{figure}

Following the methods described in the previous section, we find that the Minkowski-Bouligand dimension of the synthetic sets give expected results (\figureautorefname{ \ref{fig:4}}). It is important to note that the standard error from the linear regression model is misleading since such a model pre-supposes independence of observations and correlated data points, as in this case, underestimates it. The standard error is instead calculated from the successive differences in \(\log N_{\delta}\) as they are uncorrelated \citep{reeve_warning_1992}.

\begin{table}[b]
    \centering
    \begin{tabular}{c c c c c c c}
        \toprule
         DGGS&  \multicolumn{3}{c}{Sierpiński triangle (\(D_H \approx 1.584\))}&  \multicolumn{3}{c}{Koch curve (\(D_H \approx 1.261\))}\\
         \cmidrule(r){2-4}
         \cmidrule{5-7}
         &  \makecell{\(D_B\)}&  \makecell{Corrected \\ standard error}&  p-value&  \makecell{\(D_B\)}& \makecell{Corrected \\ standard error}& p-value\\
         \cmidrule{1-7}
         ISEA4T&1.594  &\(\pm 0.0552\)  &\(\sim 10^{-7}\)  &1.299  &\(\pm 0.0471\)  &\(0.0013\) \\
         ISEA4H&1.561  &\(\pm 0.0236\)  &\(\sim 10^{-8}\)  &1.244  &\(\pm 0.0190\)  &\(\sim 10^{-7}\) \\
         ISEA3H&1.575  &\(\pm 0.0426\)  &\(\sim 10^{-10}\)  &1.265  &\(\pm 0.0335\)  &\(\sim 10^{-5}\) \\
         \bottomrule
    \end{tabular}
    \vspace*{1mm}
    \caption{Minkowski-Bouligand dimensions (\(D_B\)) of the Sierpiński triangle and the Koch curve as measured with different DGGSs along with their standard errors and p-values}
    \label{tab:1}
\end{table}

\tableautorefname{ \ref{tab:1}} shows the computed Minkowski-Bouligand dimension, \(D_B\) of the Sierpiński triangle and the Koch curve along with the standard errors corrected for the correlation in the data points. The reported p-values are calculated from the Student t-distribution. It is observed that ISEA4T and ISEA4H DGGSs slightly overestimates and underestimates the Minkowski-Bouligand dimension respectively while values computed using ISEA3H deviates by \(<0.01\) from the theoretical Hausdorff dimension for the two sets and is the closest to the theoretical value. Aggregating the values using arithmetic mean gives \(1.576 \pm 0.0404\) and \(1.269 \pm 0.0332\) as the Minkowski-Bouligand dimensions of the Sierpiński triangle (\(D_H \approx 1.584\)) and the Koch curve (\(D_H \approx 1.261\)) respectively; i.e. with a deviation of \(0.008\) (\(0.50\%\) and \(0.63\%\) respectively).

\begin{figure}
    \centering
    \includegraphics[width=1\linewidth]{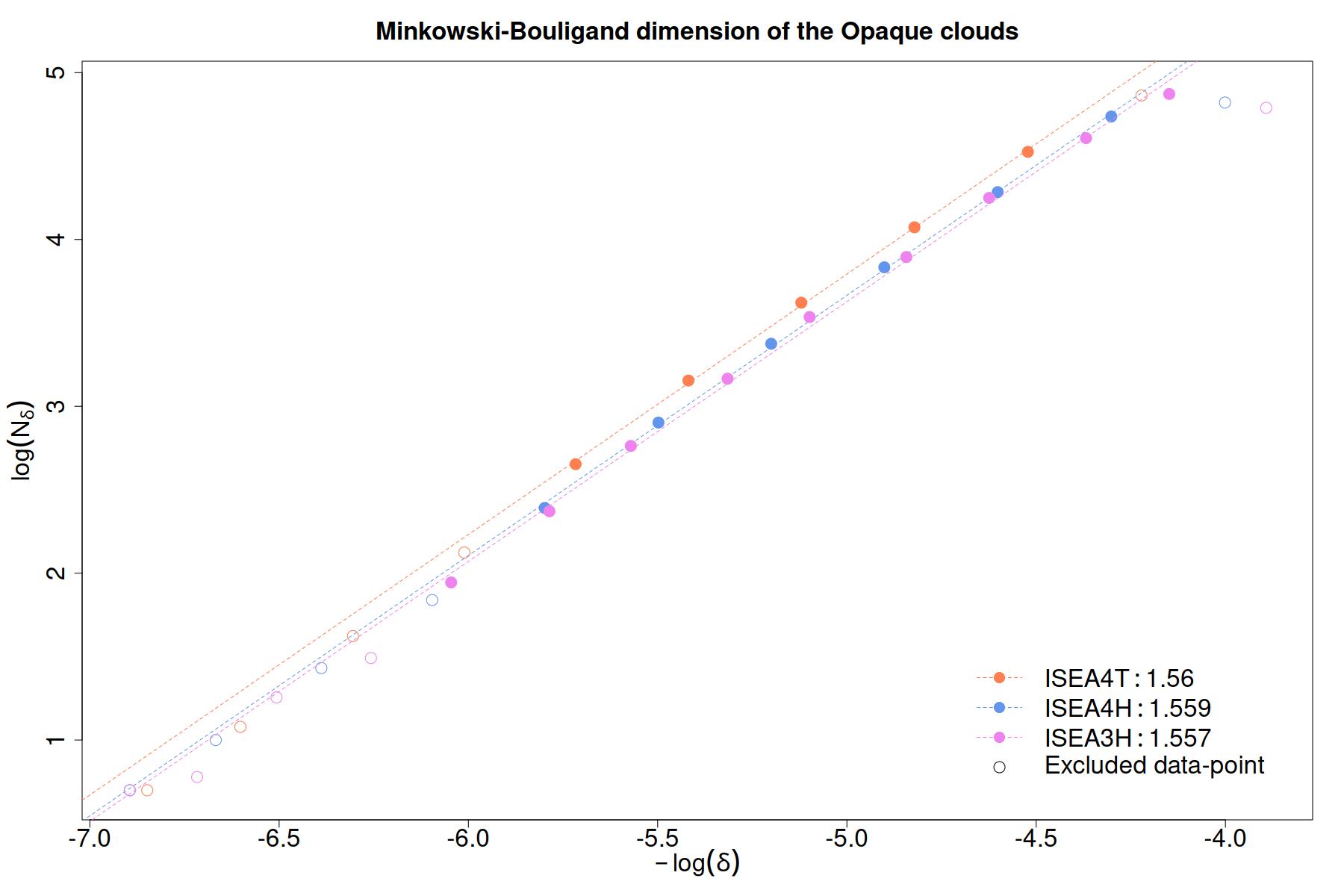}
    \caption{Calculation of the Minkowski-Bouligand dimension of the opaque cloud fields}
    \label{fig:5}
\end{figure}

\begin{table}
    \centering
    \begin{tabular}{lccc}
        \toprule
         DGGS&  \(D_B\)&  Corrected standard error& p-value\\
         \cmidrule{1-4}
         ISEA4T&  1.560&\(\pm 0.0232\)  &\(\sim 10^{-6}\) \\
         ISEA4H&  1.559&\(\pm 0.0251\)  &\(\sim 10^{-7}\) \\
         ISEA3H&  1.557&\(\pm 0.0480\)  &\(\sim 10^{-9}\) \\
         \bottomrule
    \end{tabular}
    \vspace*{1mm}
    \caption{Minkowski-Bouligand dimensions (\(D_B\)) of the opaque cloud fields as measured with different DGGSs along with their standard errors and p-values}
    \label{tab:2}
\end{table}

With strictly self-similar synthetic datasets giving expected results, we now look at a real-world opaque cloud field where the self-similarity is statistical and visually undetectable unlike our earlier examples. The Minkowski-Bouligand dimension calculated using different DGGSs (\figureautorefname{ \ref{fig:5}}) are in the range of \([1.557, 1.560]\) with a mean value of \(1.558 \pm 0.0321\), as summarised in \tableautorefname{ \ref{tab:2}}. The results obtained in this study are found to be in line with previously reported fractal dimensions of large clouds and cloud fields, as discussed in \sectionautorefname{ \ref{sec:fractal_clouds}}.

\section{Discussion}
This paper shows that the Minkowski-Bouligand dimension of geospatial feature data can be calculated practically using DGGSs as covering sets. The results obtained are in agreement with the known fractal dimensions of the respective objects. The standard errors are calculated taking into account the autocorrelation in the data used in the linear regression model.

While some deviation in the computed Minkowski-Bouligand dimension from the theoretical values has been observed, it is advisable to interpret the values with the range of their standard errors, particularly for real-world objects where self-similarity exists only in a statistical sense and not strictly as in the case of the synthetic datasets and thus calculating a single number is futile anyway. Furthermore, the synthetic datasets presented here are models based on their mathematical definitions having exactly known fractal dimensions while their computer-based representations are obtained iteratively. As the number of iterations can only have a finite value, there is bound to be some discrepancy in the form of approximation between definitions and their models that also contributes towards the deviations of the final result.

Many fractals found in the environment are multifractals -- i.e. they exhibit different fractal dimensions at different scales, locations or directions. While the study of anisotropic multifractals as such fits the framework presented here, suitable DGGSs supporting such analyses are currently unavailable. Development of DGGSs where cell shapes exhibit directional bias with increasing resolution levels would be useful for such studies.

\subsection{Choice of DGGS}
While being theoretically sound, not all DGGSs can be recommended for practical use. Given the way they are used as covering sets, two criteria emerge in this study for their use in determining Minkowski-Bouligand dimension. Performance of different DGGSs on different sets warrants a separate and thorough study but I summarise some important observances below.

\subsubsection{Low Aperture}
DGGSs of low aperture are better suited. Each resolution level of DGGSs contributes towards one data-point in the log-log plot. The more data-points there are, the better the reliability of the result. Covering any given set with a high-aperture DGGS gives fewer number of points in the region where \(\log N_{\delta} \sim \log {\delta}^{-1}\) exhibit linearity whereas for one with a lower aperture \(\delta\) increases more gradually thus obtaining more data-points. Apertures of 3 or 4 are recommended and apertures of 7 or higher are to be avoided.

Mixed aperture DGGSs with relatively low apertures such as the ISEA43H and Superfund \citep{sahr_user_2023} are also suitable. Those that use a lower aperture at higher resolutions are well-suited as capturing the variation at higher scales of measurement is more important and it provides a reasonable trade-off between number of data-points and the number of iterations required.

\subsubsection{Isodiametricity}
A less obvious aspect is the diameter of the covering sets. Literature on DGGSs mostly focus on the cells of the covering sets being of equal area as that gives several advantages for geospatial analyses. However, for the purpose of using DGGSs as covering sets the diameter is important as this is the independent variable that is varied by changing the resolution levels and represents the scale of measurement. Although the number of intersecting cells is plotted against the inverse of diameter \({\delta}^{-1}\) in a log-log scale, the value of \(\delta\) is not a single measurement but an aggregated one obtained as the supremum of the diameters of all the intersecting cells.

Consequently minimum deviation in the diameter of individual cells at each resolution level of a DGGS is a desirable property, with cells being isodiametric ideally. The presence of a cell with a large diameter with respect to the others’ is problematic as it would not be representative of the majority of the covering sets as, for the purpose of calculating the Minkowski-Bouligand dimension, the number of intersecting cells would be tabulated against this diameter of a single outlier cell. Furthermore, if this outlier occurs only at some resolution levels it is possible, particularly for mixed aperture DGGSs with large variation in apertures, that the diameter at those resolution levels is higher than that at a lower level of resolution. The method proposed here would benefit from the development of DGGSs with a focus on keeping the diameter of cells invariant at each resolution level.

\section{Conclusions}
This study links fractal dimension with DGGS by using the latter as covering sets to calculate Minkowski-Bouligand dimension. Existing literature discussing calculation of the dimension using covering sets almost exclusively use congruent square boxes embedded in the plane including for spatial objects on the earth’s surface. Consequently intersections of these boxes with the feature is performed on the plane instead of on the sphere that could give erroneous results, particularly for large geographic extents where the curvature of the earth is not trivial. In the absence of a standardised hierarchical grid, measuring box-counting dimension traditionally also includes making certain arbitrary choices guided by best practices and conventions such as the position and orientation of these boxes as well as their side-lengths that should be sampled.

Using DGGSs as covering sets instead provide standardised hierarchical grids for any point on the earth taking its curvature into account. This also solves the need for arbitrary choices in the placement of the grids since for each DGGS their cells are fixed to the earth’s surface. Calculating the Minkowski-Bouligand dimensions of synthetic datasets embedded on the earth’s surface give excellent results and a case study is presented for opaque clouds as seen from remotely sensed imagery that gives results in line with existing literature in meteorology. The validity of DGGSs as covering sets is derived from first principles and is established from a theoretical standpoint. While the method proposed is a general one applicable for use with any DGGS, use of DGGSs with low aperture and isodiametric cells is suggested for future studies adopting this method.

\section{Acknowledgements}
I thank Prof. Dr. Terence Tao (University of California, Los Angeles) for confirming that \eqref{eq:6} holds in \(\mathbb{S}^2\) \citep{tao_yes_2025}.

\bibliographystyle{apalike-ejor}

\bibliography{references}

\end{document}